\documentclass[12pt]{article}
\usepackage{amssymb,amsmath,epsfig}
\begin{document} \parindent=0pt
\parskip=6pt \rm

\begin{center}

{\bf \Large Theory of ferromagnetic superconductors with spin-triplet electron pairing}

\vspace{0.3cm}

{\bf  Dimo I. Uzunov}

 CP Laboratory, Institute of Solid State Physics, Bulgarian
Academy of Sciences, BG-1784 Sofia, Bulgaria.\\
\end{center}

{\bf Pacs}: 74.20.De, 74.25.Dw, 64.70.Tg\\

strongly correlated electrons, multi-critical point,
phase diagram.


\begin{abstract}
A general quasi-phenomenological theory that describes phases and phase transitions of ferromagnetic superconductors
with spin-triplet electron Cooper pairing is presented. The theory is based on extended Ginzburg-Landau expansion in powers
of superconducting and ferromagnetic order parameters. A simple form for the dependence of theory parameters on
the pressure ensures a correct theoretical outline of the temperature-pressure phase diagram where a stable phase of coexistence
of $p$-wave superconductivity and itinerant ferromagnetism appears. This new theory is in an excellent agreement with the
experimental data for intermetallic compounds, for example, UGe$_2$, URhGe, UCoGe, and UIr that are experimentally proven to be
itinerant ferromagnets exhibiting spin-triplet superconductivity. The mechanism of appearance of superconductivity due to
itinerant ferromagnetism ($M$-trigger effect) is established and demonstrated. On the basis of the same theory, basic features of
quantum phase transitions in this type of ferromagnetic superconductors are explained in agreement with the experimental data.
The theory allows for a classification of the spin-triplet ferromagnetic superconductors in two different types: type I and type II.
The classification is based on quantitative criteria, i.e., on simple relations between theory parameters. Both theory and experiment indicate
that the two types of $p$-wave ferromagnetic superconductors are well distinguished by essential differences in their physical properties.
\end{abstract}

\vspace{0.5cm}

{\bf 1. Introduction}

There exists an intensive research on Cooper pairing of fermions in condensed matter. Both experimental and theoretical studies reveal
fascinating phenomena of superconductivity, unconventional superfluidity and coexistence of superconductivity and ferromagnetism.
The fermion Cooper pairs in $s$-wave state are referred to as conventional pairing and, hence, the respective superfluidity or superconductivity are labeled $conventional$. The unconventional (non-$s$-wave) pairing, fore example, the $p$-wave pairing creates another type of similar phenomena -- unconventional superfluidity and unconventional superconductivity of $p$-wave type. These phenomena can be described within the general scheme of spin-triplet states
of fermion pairs. The phenomenon of unconventional (spin-triplet) Cooper pairing of fermions was firstly described for the case of superfluid Helium 3~\cite{Vollhardt:1990}. Here we focus the attention to unconventional superconductivity of spin-triplet type which allow for electron
pairing in $p$-wave states with spin equal to unity.

The spin-triplet superconducting phases are described in the framework of the general
Ginzburg-Landau (GL) effective free energy functional with the help of the symmetry
groups theory and detailed thermodynamic analysis~\cite{Volovik:1985, Blagoeva:1990};
see also~\cite{Uzunov:1990,Uzunov:2010}. This unconventional superconductivity has been discovered in
itinerant ferromagnets -- the intermetallic compounds UGe$_2$~\cite{Saxena:2000},
URhGe~\cite{Aoki:2001}, UCoGe~\cite{Huy:2007} and UIr\cite{Akazawa:2005}.

At low temperature ($T\sim 1$ K) all these compounds exhibit coexistence
of $p$-wave superconductivity and itinerant ($f$-band) electron ferromagnetism (in short, FS phase).
In the mentioned compounds, the FS phase appears only in the ferromagnetic phase domain of the $T-P$ diagram.
At given pressure $P$, the temperature
$T_{F}(P)$ of the normal-to-ferromagnetic phase (or N-FM)
transition is never lower than the temperature $T_{FS}(P)$ of the
ferromagnetic-to-FS phase (or FM-FS) transition. This is
consistent with the point of view that the superconductivity in
these compounds is triggered by the spontaneous magnetization
$\mbox{\boldmath$M$}$, as explained in details in \cite{Shopova:2003}). The
main system properties are affected by a term in the Ginzburg-Landau (GL) expansion
of the form $\mbox{\boldmath$M$}. (\mbox{\boldmath$\psi$}\times\mbox{\boldmath$\psi$})$,
which represents the interaction of $\mbox{\boldmath$M$} = \{M_j;
j=1,2,3\}$ with the complex superconducting vector field
$\mbox{\boldmath$\psi$} =\{\psi_j\}$. This term
triggers $\mbox{\boldmath$\psi$} \neq 0$ for certain $T$ and $P$
values ($\mbox{\boldmath$M$}$-trigger effect~\cite{Shopova:2003}).

The mechanism of formation of pairs in these systems is under conjectures. The electron band theory of such compounds is quite complex and the density of states $D_n(E)$ at the Fermi surface $E_F$ can hardly be evaluated in details. In this situation, the reliable theoretical interpretation of experiments is performed within the general phenomenological theory of GL type. In such theory, the Landau parameters, $\mu = (a,b,\dots)$ should depend on the shape of $D_n(E)$. While this dependence is unknown, one may reliably assume the dependence of $D_n(E)$ on temperature $T$ and pressure$P$, and hence, the double functional dependence $\mu = \mu [D_n(T,P)]$. Although unknown, the latter justifies the function $\mu(T,P)$, i.e., the dependence of the Landau parameters on $T$ and $P$. This dependence can be reliably proposed on the basis of heuristic arguments and experimental data. So, let us focus on the phenomenological approach.

The spin-triplet ferromagnetic superconductors are described by the quasi-phenomenological theory based on an extended GL in powers of the fields $\mbox{\boldmath$\psi$}$ and $\mbox{\boldmath$M$}$ (for details, see \cite{Uzunov:2010, Shopova:2003, Uzunov:2006, Uzunov:2012}). In this Report, the theory and related topics are briefly presented; for details, see~\cite{Uzunov:2010, Shopova:2003, Uzunov:2006, Uzunov:2012}. New aspects and outstanding problems are briefly indicated.

{\bf 2. Theory}

Here we hold the consideration in the lowest (mean-field) approximation. The latter is enough to elucidate the main system properties in a quite correct way. The free energy per unit volume, $F/V =f(\mbox{\boldmath$\psi$},\mbox{\boldmath$M$})$, can
be written in the form
\begin{eqnarray}
\label{Eq1}f(\mbox{\boldmath$\psi$},\mbox{\boldmath$M$}) & = & a_s|\mbox{\boldmath$\psi$}|^2 +\frac{b_s}{2}|\mbox{\boldmath$\psi$}|^4 +
  \frac{u_s}{2}|\mbox{\boldmath$\psi$}^2|^2 +\frac{v_s}{2}\sum_{j=1}^{3}|\psi_j|^4\\ \nonumber
 && +
 a_f\mbox{\boldmath$M$}^2 +
 \frac{b_f}{2}\mbox{\boldmath$M$}^4 + i\gamma_0 \mbox{\boldmath$M$}
 \cdot (\mbox{\boldmath$\psi$}\times \mbox{\boldmath$\psi$}^*) + \delta
\mbox{\boldmath$M$}^2 |\mbox{\boldmath$\psi$}|^2.
\end{eqnarray}
\noindent The material parameters satisfy $b_s >0$, $b_f>0$, $a_s = \alpha_s(T-T_{s})$,
and $a_f = \alpha_{f}[T-T_{f}(P)]$. The terms proportional to $u_s$ and $v_s$
describe, respectively, the anisotropy of the spin-triplet
electron Cooper pairs and the crystal anisotropy. Next, $\gamma_0
\sim J$ (with $J>0$ the ferromagnetic exchange constant) and
$\delta > 0$ are parameters of the
$\mbox{\boldmath$\psi$}$-$\mbox{\boldmath$M$}$ interaction terms.
Previous mean-field studies~\cite{Shopova:2003} have shown that the anisotropy
represented by the $u_s$ and $v_s$ terms in Eq. (\ref{Eq1})
slightly perturb the size and shape of the stability domains of
the phases, while similar effects can be achieved by varying the
$b_s$ factor in the $b_s|\mbox{\boldmath$\psi$}|^4$ term. For
these reasons, in the present analysis we ignore the anisotropy
terms, setting $u_s = v_s = 0$, and consider $b_s\equiv b >0$ as
an effective parameter. Then, without loss of generality, we use the gauge
$\mbox{\boldmath$M$} = (0,0,M)$.

A convenient dimensionless free energy can now be defined by
$\tilde{f} = f/(b_f M_0^4)$, where $M_0 = [\alpha_fT_{f0}
/b_f]^{1/2} >0$ is the value of $M$ corresponding to the pure
magnetic subsystem $(\mbox{\boldmath$\psi$} \equiv 0)$ at $T=P=0$
and $T_{f0}=T_f(0)$. On scaling the order parameters as $m =
M/M_0$ and $\mbox{\boldmath$\varphi$} = \mbox{\boldmath$\psi$}
/[(b_f/b)^{1/4}M_0]$ we obtain

\begin{equation}
\label{Eq2} \tilde{f}= r\phi^2 + \frac{\phi^4}{2}+ tm^2
+\frac{m^4}{2} + 2\gamma m\phi_1\phi_2\mbox{sin}\theta +
\gamma_1m^2\phi^2,
\end{equation}

\noindent where $\phi_j =|\varphi_j|$, $\phi =
|\mbox{\boldmath$\varphi$}|$, and $\theta$ is the phase angle
between the complex $\varphi_2$ and $\varphi_1$. The dimensionless
constants are $t = [\tilde{T}-\tilde{T}_f(P)]$, $r = \kappa
(\tilde{T}-\tilde{T}_s)$ with $\kappa =
\alpha_sb_f^{1/2}/\alpha_fb^{1/2}$, $\gamma =
\gamma_0/ [\alpha_fT_{f0}b]^{1/2}$, and $\gamma_1 =
\delta/(bb_f)^{1/2}$. The reduced temperatures are $\tilde{T} =
T/T_{f0}$, $\tilde{T}_f(P) = T_f(P)/T_{f0}$, $\tilde{T}_s(P)=
T_s(P)/T_{f0}$.

\begin{figure}
\begin{center}
\epsfig{file=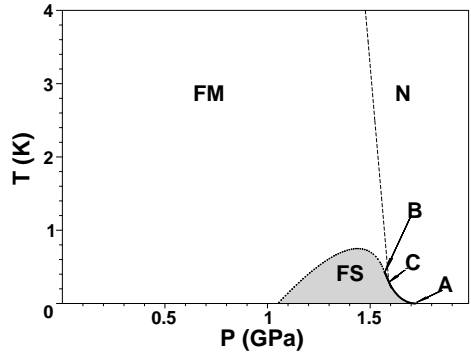,angle=-90, width=6cm}
\end{center}
\caption{\small Low-temperature part of the $T-P$ phase
diagram of UGe$_2$, shown in Fig.~1. The points A, B, C are
located in the high-pressure part ($P\sim P_c\sim 1.6$ GPa). The
FS phase domain is shaded. The solid lines AC and BC show
the first-order transitions N-FS, and FM-FS, respectively. The dotted line shows a second-order FS-FM transition,
the dashed line stands for the N-FM transition.}
\end{figure}

{\bf 3. Results}

The analysis involves making simple assumptions for the $P$
dependence of the $t$, $r$, $\gamma$, and $\gamma_1$ parameters in
Eq. (\ref{Eq2}). Specifically, we assume that only $T_f$ has a
significant $P$ dependence, described by $\tilde{T}_f(P) = (1 -
\tilde{P})$, where $\tilde{P} = P/P_0$ and $P_0$ is a
characteristic pressure deduced later; for UGe$_2$, $P_0 \sim P_c$ - the critical pressure
above which both magnetic and supercionducting order
vanish at $T \sim 0$ (see Fig.~1); for some substances this is not the case,
but then another dependence of $T_f$ on $P$ takes place. Within this simple variant,
all other theory parameters are considered $P$-independent.

The simplified model in Eq. (\ref{Eq2}) is capable of describing
the main thermodynamic properties of spin-triplet ferromagnetic
superconductors\cite{Shopova:2003}. For example, in Fig.~1 the low temperature part of the $T-P$ diagrams of UGe$_2$ is shown. The result is in a remarkable agreement with the experimental shape of the same diagram~\cite{Saxena:2000}. The type of the phase transitions and the multicritical points (A, B, C) also correspond to the experimental observations. There are three stable phases: (i) the normal (N)
phase, given by $\phi = m = 0$; (ii) the pure ferromagnetic (FM)
phase, given by $m = (-t)^{1/2} > 0$, $\phi =0$; and (iii) the FS
phase, given by $\phi_1^2= \phi_2^2= ( \gamma m-r-\gamma_1
m^2)/2$, $\phi_3 = 0$, where $\mbox{sin}\theta = -1$ and $m$
satisfies

\begin{equation}
\label{Eq3} (1-\gamma_1^2) m^3 + \frac{3}{2} \gamma \gamma_1 m^2
+\left(t-\frac{\gamma^2}{2}-\gamma_1 r\right) m + \frac{\gamma
r}{2}=0.
\end{equation}

\noindent Although Eq. (\ref{Eq3})
is complicated, some analytical results follow, e.g., we find that
the second order phase transition line $\tilde{T}_{FS}(P)$
separating the FM and FS phases (the dotted lines in Figs.~1) is the solution of

\begin{equation}
\label{Eq4} \tilde{T}_{FS}(P) = \tilde{T}_s +
\frac{\gamma_1}{\kappa}t(T_{FS}) +
\frac{\gamma}{\kappa}[-t(T_{FS})]^{1/2}.
\end{equation}

Under certain conditions, which are satisfied for UGe$_2$, the $T_{FS}(P)$ curve has a maximum at
$\tilde{T}_{m} = \tilde{T}_s +(\gamma^2/4\kappa\gamma_1)$ with
pressure $P_m$ found by solving $t(T_m,P_m)=
-(\gamma^2/4\gamma_1^2)$.


Under certain conditions, Eq.~(2) describes pure superconducting phases, too.
To date there are no experimental observation of such phase,
which means that the respective conditions are not satisfied by the parameters
of the currently known superconductors.  Negative values of $T_s$ are possible, and they
describe a phase diagram topology in which the FM-FS transition
line terminates at $T=0$ for $P< P_c$. This might be of relevance
for other compounds, e.g., URhGe.

As in experiment\cite{Saxena:2000}, the dashed line in Fig.~1 slopes to $T=0$ at another critical pressure $P_{0c}$.
This is possible in UGe$_2$ because for this compound the condition $\gamma < \gamma_1$ is satisfied. Such systems
are spin-triplet ferromagnetic superconductors of type I, whereas systems which fulfill the condition
$\gamma > \gamma_1$ are of type II \cite{Shopova:2003, Uzunov:2012}. The quantum phase transitions at
$P_0c$ and $P_c$ have remarkable properties, as shown in \cite{Uzunov:2006}. Depending on the system properties,
$T_C$ -- the temperature locating point $C$, can be either positive
(when a direct N-FS first order transition is possible), zero, or
negative (when the FM-FS and N-FM phase transition lines terminate
at different zero-temperature phase transition points). The last
two cases correspond to $T_s < 0$. All these cases are possible in
spin-triplet ferromagnetic superconductors. The zero temperature transition at
$P_{c0}$ is found to be a quantum critical point, whereas the
zero-temperature phase transition at $P_c$ is of first order. As
noted, the latter splits into two first order phase transitions. This classical
picture may be changed through quantum fluctuations
\cite{Shopova:2003}.

The quantum phase transitions at
$P_{0c}$ and $P_c$ have remarkable properties. An investigation \cite{Uzunov:2006} performed
by renormalization group methods revealed a
fluctuation change in the order of the zero temperature first order phase transition at $P_c$ to a continuous
phase transition belonging to an entirely new class of
universality. However, this option exists only for magnetically
isotropic order (Heisenberg symmetry) and is unlikely to apply in
the known spin-triplet ferromagnetic superconductors, which are magnetically anisotropic.

The application of the theory was demonstrated on the example of UGe$_2$, but the same theory has ample volume of options to describe various
real superconductors of the same type and is not restricted to a particular compound, or, to a particular group of such materials.

Even in its simplified form, this theory has been shown to be
capable of accounting for a wide variety of experimental behavior.
A natural extension to the theory is to add a
$\mbox{\boldmath$M$}^6$ term which provides a formalism to
investigate possible metamagnetic phase transitions and extend some first order phase transition
lines, as required by experimental data. Another modification of this theory, with regard to
applications to other compounds, is to include a $P$ dependence
for some of the other GL parameters.

Among the outstanding problems are: local gauge effects on the vortex phase and the phase transitions, the outline of the upper critical magnetic field $H_{c2}(T,P)$, thermal and magnetic properties, and the description of $T-P$ diagrams with topologies, which are observed in experiments. Such studies require an extension of the theory by including $\mbox{\boldmath$M$}^6$ term, the magnetic induction $\mbox{\boldmath$B$} = \mbox{\boldmath$H$} + 4\pi \mbox{\boldmath$M$}$, and a more precise $P$-dependence of some theory parameters.

{\bf Acknowledgements:} The author is grateful to A. Harada and
S. M. Hayden for valuable discussions of experiments.


\begin{thebibliography}{ll}
\bibitem{Vollhardt:1990}
D. Vollhardt and P. W\"olfle (1990) {\em The Superfluid Phases of
Helium 3}, Taylor $\&$ Francis, London.
\bibitem{Volovik:1985}
G. E. Volovik and L. P. Gor'kov (1985) {\em Sov. Phys. JETP} {\bf
61} 843 [{\em Zh. Eksp. Teor. Fiz.} {\bf 88} (1985) 1412].
\bibitem{Blagoeva:1990}
E. J. Blagoeva, G. Busiello, L. De Cesare, Y. T. Millev, I.
Rabuffo, and D. I. Uzunov (1990) {\em Phys. Rev.} {\bf B42} 6124.
\bibitem{Uzunov:1990}
D. I. Uzunov (1990) in {\em Advances in Theoretical Physics},
 ed. by E. Caianiello; World Scientific, Singapore, p. 96.
\bibitem{Uzunov:2010}
D. I. Uzunov (2010) {\em Theory of Critical Phenomena}, World
Scientific, Singapore (Second edition); First edition (1993).
\bibitem{Saxena:2000}
S. S. Saxena, P. Agarwal, K. Ahilan, F. M. Grosche, R. K. W.
Haselwimmer, M.J. Steiner, E. Pugh, I. R. Walker, S.R. Julian, P.
Monthoux, G. G. Lonzarich, A. Huxley. I. Sheikin, D. Braithwaite,
and J. Flouquet (2000)  {\em Nature} {\bf 406} 587.
\bibitem{Aoki:2001}
D. Aoki, A. Huxley, E. Ressouche, D. Braithwaite, J. Flouquet,
J-P.. Brison, E. Lhotel, and C. Paulsen (2001) {\em Nature} {\bf
413} 613.
\bibitem{Huy:2007}
N. T. Huy, A. Gasparini, D. E. de Nijs, Y. Huang, J. C. P.
Klaasse, T. Gortenmulder, A. de Visser, A. Hamann, T. G\"orlach,
and H. v. L\"ohneysen(2007) Phys. Rev. Lett. {\bf 99} 067006 .
\bibitem{Akazawa:2005}
T. Akazawa, H. Hidaka, H. Kotegawa, T. C. Kobayashi, T. Fujiwara,
E. Yamamoto, Y. Haga, R. Settai, and Y. Onuki (2005) Physica B {\bf
359-361} 1138.
\bibitem{Shopova:2003}
D. V. Shopova and D. I. Uzunov (2003) {\em Phys. Lett. A} {\bf
313} 139; D. V. Shopova and D. I. Uzunov (2005) {\em Phys. Rev. B} {\bf 72} 024531;
D. V. Shopova and D. I. Uzunov (2009) {\em Phys. Rev. B} {\bf 79} 064501.
\bibitem{Uzunov:2006}
 D. I. Uzunov (2006) {\em Phys. Rev. B} {\bf 74} 134514; D. I. Uzunov (2007) {\em Europhys. Lett.} {\bf 77} 20008.
\bibitem{Uzunov:2012}
 D. I. Uzunov (2012) in {\em Superconductors}, ed. by A. Gabovich; Intech, Rijeka, Ch. 17, pp 415-440.
\end{thebibliography}
\end{document}